\documentclass[pra,twocolumn,superscriptaddress,amsmath,amssymb,floatfix,noshowpacs]{revtex4-2} 
\usepackage{graphicx}   
\usepackage{dcolumn}    
\usepackage{bm}         
\usepackage{hyperref}   
\usepackage[toc,page]{appendix}

\begin{document}

\title{\textbf{Dissipative transfer of quantum correlations from light to atomic arrays}}
\author{Roni Ben-Maimon}
\affiliation{Department of Chemical \& Biological Physics, Weizmann Institute of Science, Rehovot 7610001, Israel}
\author{Yakov Solomons}
\affiliation{Department of Chemical \& Biological Physics, Weizmann Institute of Science, Rehovot 7610001, Israel}
\author{Ephraim Shahmoon}
\affiliation{Department of Chemical \& Biological Physics, Weizmann Institute of Science, Rehovot 7610001, Israel}
\date{\today}

\begin{abstract}
We show how the directional collective response of atomic arrays to light can be exploited for the dissipative generation of entangled atomic states, relevant for e.g. quantum metrology. We consider an atomic array illuminated by a paraxial beam of a squeezed-vacuum field and demonstrate that quantum-squeezing correlations are dissipatively transferred to the array atoms, resulting in an atomic spin-squeezed steady state. We find that the entanglement transfer efficiency and hence the degree of spin squeezing are determined by the resonant optical reflectivity of the array. Considering realistic cases of finite-size array and illuminating beam, we find how the spin-squeezing strength scales with system parameters, such as the number of layers in the array and its spatial overlap with the beam. We discuss applications in atomic clocks both in optical and microwave domains.
\end{abstract}

\maketitle


Quantum-correlated states of atoms and spins are of high relevance for quantum information processing and quantum-enhanced metrology and sensing \cite{ref01,ref02,ref03,ref04,ref38}. One of the most valuable forms of quantum correlations and entanglement for quantum metrology is that of a spin-squeezed state, exhibiting reduced quantum projection noise below the standard quantum limit, thus allowing to achieve quantum-enhanced measurement sensitivities \cite{ref05,ref06,ref07}. Such a quantum enhancement is crucial for further development of atomic clocks, which already operate near the standard quantum limit \cite{ref32,ref33,ref30}. In particular, this applies to state-of-the-art systems comprised of arrays of trapped atoms operating as optical clocks, as in optical lattice clocks and tweezer-array clocks \cite{ref30,ref31,ref44,ref45}, motivating the need to produce spin squeezing in atomic ensembles and specifically in atomic arrays.

A prevalent approach for the generation of spin squeezing in atomic systems is to design effective dispersive or dissipative interactions between the atoms. This can be achieved using techniques ranging from the quantum non-demolition measurement of a common optical mode interacting with all atoms \cite{ref21,ref22,ref46,ref47,ref42}, to engineering mediated spin-spin interactions leading to Hamiltonian one-axis or two-axis twisting dynamics \cite{ref09,ref29,ref43,ref31,ref48,ref45,POH2}, or designing collective radiation and driven-dissipative dynamics to generate squeezing \cite{ref24,ref50,ref51,ref52,ref53,ref54}.

\begin{figure}
  \centering
  \includegraphics[width=\columnwidth]{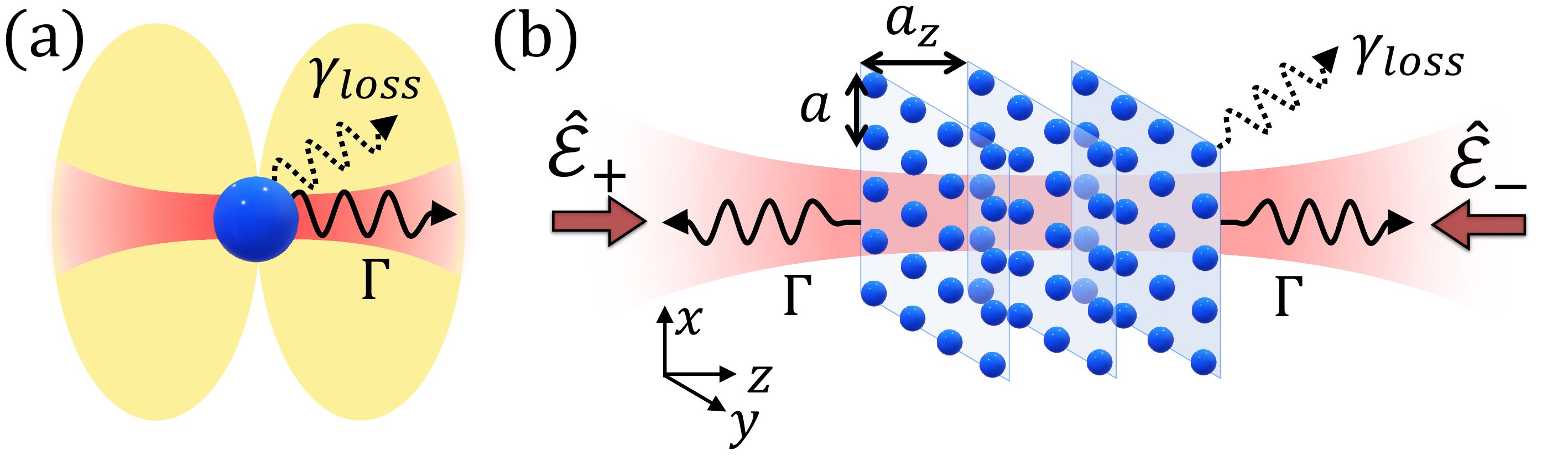}
  \caption{Coupling of correlated paraxial light to atoms. (a) The individual-atom dipole radiation pattern (yellow shaded region) matches predominantly to the vacuum state of non-paraxial modes (rate $\gamma_{\mathrm{loss}}$ close to the natural decay rate $\gamma$) and only weakly couples to paraxial modes occupied by correlated light (rate $\Gamma\ll \gamma$, red shaded region). (b) An atom array comprised of multiple layers of subwavelength  square lattices is illuminated by correlated squeezed light. The directional collective coupling of the array to light yields strong coupling to the paraxial correlated light [$\Gamma=\mathcal{O}(\gamma)$] and highly suppressed coupling to undesired vacuum modes ($\gamma_{\mathrm{loss}}\ll \gamma$).}\label{fig1}
\end{figure}

An alternative approach is to employ available light sources which exhibit quantum squeezed fluctuations \cite{ref34,ref35,ref36,ref37,ref41} and to transfer their squeezing correlations to the atoms either dissipatively \cite{ref11,ref12} or coherently \cite{YEL1,ref13,ref14}. Such an approach requires a strong coupling between the squeezed light and the atomic system \cite{ref10,ref23,ref55}, which is typically hindered by the poor spatial overlap between paraxial squeezed-light sources and the dipole radiation pattern of an atom in free space (Fig. \ref{fig1}a). Initial methods to achieve dissipative squeezing transfer in free-space atomic ensembles were restricted by a conversion ratio of $1/2$ even for a large optical depth \cite{ref11,ref12}, whereas subsequent proposals involving spatially confined photon modes in cavities or waveguides may entail technical challenges, such as mode matching between the squeezed light and the confined mode, and atom trapping near nanostructures \cite{ref25,ref26,ref27}.

\begin{figure}
  \centering
  \includegraphics[width=\columnwidth]{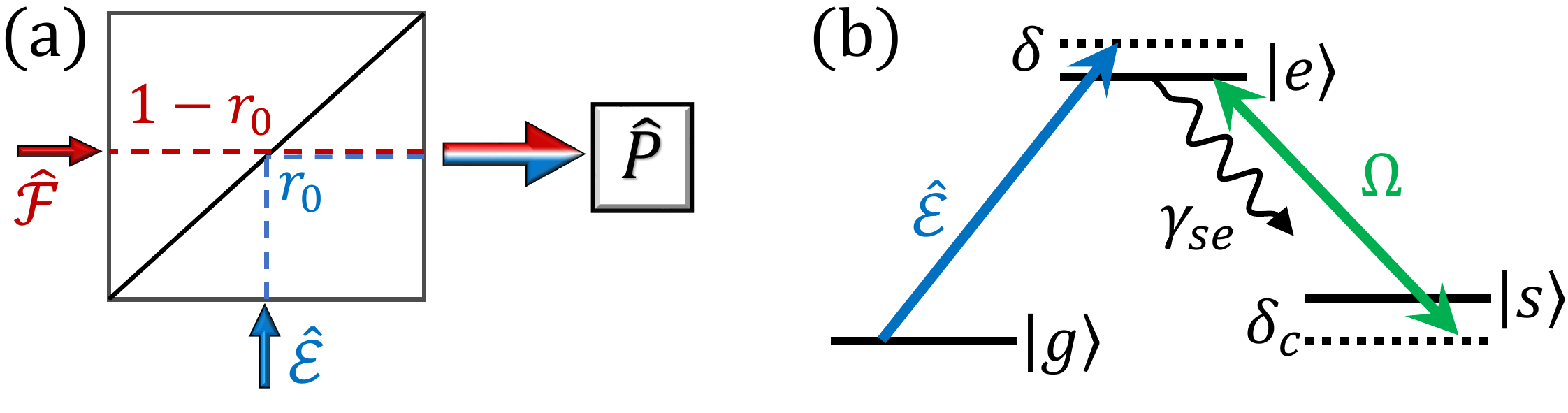}
  \caption{(a) Beam splitter description of Eq. (\ref{eq_Pu_HL}). The atomic array is fed by two sources: quantum-correlated input-light $\hat{{\cal E}}$ with coupling strength given by the array reflectivity $r_{0}$, and vacuum noise $\hat{{\cal F}}$ coupled through $1-r_{0}$. (b) 3LA scheme. Interaction between the field $\hat{{\cal E}}$ and the collective coherence of the levels $|g\rangle$ and $|s\rangle$ can be switched off by setting $\Omega=0$.}\label{fig2}
\end{figure}

Here we show, however, that the dissipative transfer of quantum correlations from light to atoms is in fact very natural when dealing with ordered atomic arrays, providing a promising approach for robust, steady-state spin squeezing generation in e.g. lattice clocks. The idea is that the combination of collective optical response of the atoms, formed by their induced dipolar interactions, together with the array spatial order, leads to a strong and directional coupling to light \cite{ref19,ref16,ref15,ref17,ref18,MAZ,ZOL,POH,ref28,YEL,ALJ,PET,ref20}. The strength of this coupling is evident by the high reflectivity of the array, which characterizes the efficiency of various quantum tasks \cite{ref15,ref16,ref17,ref20,ref28}. Utilizing this property, we demonstrate that the illumination of an atom array with a paraxial squeezed-vacuum field leads to a spin-squeezed steady state of the atoms, and that the fraction of quantum squeezing transferred from the light is quantified by the array reflectivity. We obtain these results by mapping a realistic many-atom problem of two- or three-level atoms, onto a simple beam-splitter model characterized by the array reflectivity (Fig. \ref{fig2}a). This allows us to obtain analytical results for the scaling of the achievable spin squeezing with system resources, such as the number of array layers and the parameters of the squeezed light source.

\emph{The system}--- We consider a three-dimensional (3D) array of identical atoms illuminated from both sides by paraxial correlated light at a carrier wavelength $\lambda$, later taken as a squeezed vacuum field (Fig. \ref{fig1}b). The array consists of $N_{z}$ layers along the $z$ axis with interlayer distance $a_z$, each layer forming a 2D square lattice in the $xy$ plane with a subwavelength lattice constant $a<\lambda$. The corresponding atomic positions are $\boldsymbol{r}_{\boldsymbol{n}_{\perp}n_{z}}=(\boldsymbol{n}_{\perp}a,n_z a_z)$ with indices $\boldsymbol{n}_{\bot}=(n_x,n_y)$, $n_{x},n_{y}\in\left\{ 0,1,\dots N-1\right\}$ and $n_{z}\in\left\{ 0,1,\dots N_{z}-1\right\}$. First, we consider two-level atoms (TLAs) and subsequently expand the model to incorporate three-level atoms (3LAs), relevant to optical and microwave atomic clocks, respectively. Under the Born-Markov approximation, we arrive at the Heisenberg-Langevin equation for the lowering atomic operators, $\hat{\sigma}_{\boldsymbol{n}_{\perp}n_{z}}=\left(\left|g\right\rangle \left\langle e\right|\right)_{\boldsymbol{n}_{\perp}n_{z}}$, within the weak driving linear regime wherein most atoms are in their ground state,
\begin{eqnarray}
\label{eq_HL_position}
\begin{aligned}
        \displaystyle
\dot{\hat{\sigma}}_{\boldsymbol{n}_{\perp}n_{z}}{}&=  \left(i\delta-\frac{\gamma_{s}}{2}\right)\hat{\sigma}_{\boldsymbol{n_{\perp}}n_{z}}+\hat{F}_{\boldsymbol{n}_{\perp}n_{z}}+i\frac{d}{\hbar}\hat{E}_{0}\left(\boldsymbol{r}_{\boldsymbol{n}_{\perp}n_{z}}\right)\\
&+i\frac{3}{2}\gamma\lambda\sum_{\boldsymbol{m}_{\perp},m_{z}}G\left(\omega,\boldsymbol{r}_{\boldsymbol{m}_{\perp}m_{z}}-\boldsymbol{r}_{\boldsymbol{n}_{\perp}n_{z}}\right)\hat{\sigma}_{\boldsymbol{m}_{\perp}m_{z}}.
\end{aligned}
\end{eqnarray}
Here, $\delta=\omega-\omega_{e}$ is the detuning between the central frequency of the incident field, $\omega=2\pi c/\lambda$, and the atomic transition, $\omega_{e}$, while $\gamma_{s}$ and the associated quantum Langevin noise $\hat{F}_{\boldsymbol{n}_{\perp}n_{z}}$ account for single-atom, non-collective decay processes due to imperfections such as position disorder or other decay channels \cite{ref20}. The incident, freely-evolving field  is described by the operator $\hat{E}_{0}(\boldsymbol{r})$, already projected onto the atomic dipole orientation, with $d$ the dipole matrix element. Collective response of the array is established by the photon-mediated dipole-dipole interactions between the atoms, accounted for by the dyadic Green’s function of the photons, $G\left(\omega,\boldsymbol{r}_{\boldsymbol{m}_{\perp}m_{z}}-\boldsymbol{r}_{\boldsymbol{n}_{\perp}n_{z}}\right)$, evaluated at frequency $\omega$ and projected onto the dipole orientation.

For an ``ideal" infinite array, the discrete translational symmetry guarantees in-plane momentum conservation and hence directional coupling between momentum-space collective-dipole eigenmodes of the array and corresponding plane waves of light \cite{ref15}. Although a realistic, finite-size array does not possess translational symmetry, it is shown in Ref. \cite{ref20} that for paraxial illumination directional coupling to light is still maintained: in this case the spatially matched collective dipole mode of the atoms forms an excellent approximation of a dipole eigenmode, allowing for a mapping to a simple 1D-like problem of scattering. To this end, we consider a paraxial input-field mode propagating along the $\pm z$ direction, with a transverse spatial profile $u\left(\boldsymbol{r}_{\perp}\right)$, satisfying $\int d\boldsymbol{r}_{\perp}\left|u\left(\boldsymbol{r}_{\perp}\right)\right|^{2}=1$, e.g. a Gaussian beam of width $w$, $u\left(\boldsymbol{r}_{\perp}\right)=\sqrt{\frac{2}{\pi w^{2}}}e^{-\left|\boldsymbol{r}_{\perp}\right|^{2}/w^{2}}$. We define the projection of the electric field onto the paraxial mode, and the corresponding, spatially matched collective dipole $\hat{P}$ as
\begin{eqnarray}\label{Eu_transformation}
&&\hat{{\cal E}}_{\pm}\left(z\right)=\frac{d}{\hbar}\sqrt{\frac{2}{a^{2}\Gamma_{0}}}\intop d^{2}\boldsymbol{r}_{\perp}\hat{E}_{0,\pm}\left(\boldsymbol{r}_{\perp},z\right)u\left(\boldsymbol{r}_{\perp}\right),
\nonumber\\
&&\hat{P}=\frac{1}{\sqrt{N_{z}}}\frac{a}{\sqrt{\eta}}\sum_{\boldsymbol{n}_{\perp},n_{z}}
e^{i \frac{2\pi }{\lambda}a_z n_{z}}u\left(\boldsymbol{r}_{\boldsymbol{n}_{\perp}}\right)\hat{\sigma}_{\boldsymbol{n_{\perp}}n_{z}}.
\label{trans}
\end{eqnarray}
Here, $\hat{{\cal E}}_{\pm}\left(z\right)$ [and $\hat{E}_{0,\pm}(\boldsymbol{r})$] are right- ($+$) and left- ($-$) propagating fields (Appendix A), $\Gamma_{0}$ is defined in Eq. (\ref{eq_emission_rates}) below, and $\eta=\text{erf}^{2}\left(\frac{Na}{\sqrt{2}w}\right)$ is the overlap between the beam and the array, yielding the commutation relations $[\hat{{\cal E}}_{\pm}(z),\hat{{\cal E}}_{\pm}^{\dagger}(z')]=\delta(z-z')$ and $[\hat{P},\hat{P}^{\dagger}]=1$ (in the linear regime where $\hat{P}$ is a boson).

Following Ref. \cite{ref20}, we consider conditions which guarantee directional coupling between $\hat{P}$ and $\hat{{\cal E}}_{\pm}$: (i) The linear size of each 2D layer, $Na$, is assumed larger than the dipole-dipole interaction lengthscale, $\lambda$, so that the majority of the atoms experience infinite array-like interactions. (ii) The on-axis length of the array, $\left(N_{z} - 1\right) a_{z}$ is assumed smaller than the Rayleigh distance of the input light, $z_R=\pi w^2/\lambda$, such that the latter exhibits 1D-like propagation along the array. (iii) The interlayer spacing is taken as an integer multiple of a wavelength, $a_{z}/\lambda\in\mathbb{N}$, leading to phase-matching conditions between interfering layers identical to both backward and forward incident light. (iv) We assume that the interlayer spacing satisfies $a_z\gtrsim a$, making the contribution of evanescent fields to the dipole-dipole interaction between the layers negligible.

\emph{Equivalent beam-splitter description}--- Implementing the transformation (\ref{trans}) on Eq. (\ref{eq_HL_position}), and considering the conditions presented above, we obtain a dynamical equation for $\hat{P}$ and its coupling to the symmetric field, $\hat{{\cal E}}(t)=[\hat{{\cal E}}_{+}(z=0,t)+\hat{{\cal E}}_{-}(z=0,t)]/\sqrt{2}$ \cite{ref20},
\begin{eqnarray}\label{eq_Pu_HL}
\begin{aligned}
        \displaystyle
\dot{\hat{P}}&=\left[i\left(\delta-\Delta\right)-\frac{\Gamma+\gamma_{\text{loss}}}{2}\right]\hat{P}+i\sqrt{\Gamma}\hat{{\cal E}}(t)+i\sqrt{\gamma_{\text{loss}}}\hat{{\cal F}}(t),
\end{aligned}
\end{eqnarray}
with,
\begin{eqnarray}\label{eq_emission_rates}
\begin{aligned}
\Gamma=\eta N_{z}\Gamma_{0},\ \gamma_{\text{loss}}=\left(1-\eta\right)\Gamma_{0}N_{z}+\gamma_{s},\
&\Gamma_{0}=\frac{3\gamma}{4\pi}\frac{\lambda^{2}}{a^{2}}.
\end{aligned}
\end{eqnarray}
Here, $\Gamma$ is the collective damping rate to the input-field mode, $\hat{{\cal E}}(t)$, and $\gamma_{\text{loss}}$ describes the emission to undesired modes, where $\Gamma_{0}$ and $\Delta$ are the cooperative emission rate and energy shift of an infinite 2D array \cite{ref15}. The quantum noise $\hat{{\cal F}}(t)$ describes vacuum fluctuations corresponding to $\gamma_{\text{loss}}$ and is originated in two sources, $i\sqrt{\gamma_{\text{loss}}}\hat{{\cal F}}(t)=\hat{F}_{u}+\hat{F}_{\eta}$, with $\hat{F}_{u}$ being the transformation of the noise $\hat{F}_{\boldsymbol{n}_{\perp}n_{z}}$, and $\hat{F}_{\eta}$ pertaining to vacuum noise resulting from coupling to undesired leakage modes due to the finite size of the array, corresponding to $\left(1-\eta\right)\Gamma_{0}N_{z}$.

On resonance, Eq. (\ref{eq_Pu_HL}) reveals a simple beam splitter interaction between light and atoms, characterized by the on-resonance reflectivity of the array \cite{ref15,ref20},
\begin{eqnarray}\label{eq_r0}
r_0=\frac{\Gamma}{\Gamma+\gamma_{\text{loss}}}.
\label{r0}
\end{eqnarray}
As seen in Fig. \ref{fig2}a, the linearized bosonic mode $\hat{P}$ of the array is fed by two ports: its coupling strength to the desired input-mode of correlated light is given by $r_0$, while the coupling to the vacuum noise inserted by lossy modes is $1-r_{0}$, thus $r_0$ is expected to quantify the efficiency of correlation transfer from light to matter.

\emph{Squeezing of light and atoms}---
We now consider the concrete case of an input field in a squeezed vacuum state, characterized by the correlations \cite{ref39} (Appendix A),
\begin{eqnarray}\label{eq_squeezed_correlations}
\begin{aligned}
       \displaystyle
\langle \hat{{\cal E}}^{\dagger}(t)\hat{{\cal E}}(t')\rangle =\mathcal{N}\delta\left(t-t'\right),\
& \langle \hat{{\cal E}}(t)\hat{{\cal E}}(t')\rangle =\mathcal{M}\delta\left(t-t'\right),
\end{aligned}
\end{eqnarray}
Here $\mathcal{N}$ is the average photon number, $\mathcal{M}$ is taken real without loss of generality, and both parameters are related by $\mathcal{M}=\alpha\sqrt{\mathcal{N}\left(\mathcal{N}+1\right)}$, where $0\leqslant\alpha\leqslant 1$ represents the purity of a practical squeezed-vacuum source ($\alpha=1$ describing pure entangled photon pairs). The field exhibits quantum squeezing when its minimal quadrature noise, given by $\xi_{F}^{2}=1+2(\mathcal{N}-\mathcal{M})$, drops below $1$ (below the vacuum noise).

Squeezing correlations are expected to be transferred to the atoms in the form of spin squeezing. As our system exhibits non-uniform coupling to light, spin squeezing is defined for a collective spin operator, $\hat{\boldsymbol{L}}$, whose spatial distribution matches that of the input field \cite{ref40,ref42,ref43}. We define the spin squeezing parameter as the ratio of phase sensitivities in Ramsey measurements using either a squeezed spin state or a coherent spin state (CSS) \cite{ref06,ref07}, $\xi^{2}=\text{Var}(\phi)/\text{Var}(\phi_{\text{CSS}})$, where the phase sensitivities are evaluated for the spatially matched spin operator $\hat{\boldsymbol{L}}$, $\text{Var}\left(\phi\right)=\min_{\theta}\text{Var}(\frac{1}{2}\hat{L}_{\theta})/|\langle\hat{\boldsymbol{L}}\rangle|^{2}$,
with $\hat{L}_{\theta}=\hat{L}_+e^{-i\theta}+\hat{L}_-e^{i\theta}$ and we take $|\langle \hat{\mathbf{L}} \rangle|\approx |\langle \hat{L}_z \rangle|$ in the linearized regime. The spatially matched lowering operator $\hat{L}_-=\hat{L}_x-i\hat{L}_y$ is a weighted sum of individual atomic lowering operators with the corresponding weights proportional to the non-uniform coupling to the field, $\hat{L}_-\propto\sum_{\boldsymbol{n}_{\perp}n_{z}}e^{ik_{p}a_{z}n_{z}}u\left(\boldsymbol{r}_{\boldsymbol{n}_{\perp}}\right)\hat{\sigma}_{\boldsymbol{n_{\perp}},n_{z}}$, and a normalization factor that ensures that the components of $\hat{\boldsymbol{L}}$ satisfy on average the SU(2) commutation relations \cite{ref40} (Appendix B). Therefore, $\hat{L}_-$ is in fact proportional to the collective dipole $\hat{P}$ defined above, and in the considered linear regime, the spin squeezing parameter becomes identical to the squeezing parameter of a boson $\hat{P}$,
\begin{eqnarray}
        \displaystyle
\xi^{2}=\min_{\theta}\text{Var}\left(\hat{P}e^{i\theta}+\hat{P}^{\dagger}e^{-i\theta}\right).
\label{eq_squeezing_param_def}
\end{eqnarray}
Spin squeezing exists for $\xi^{2}<1$ and serves as an entanglement witness of the collective atomic state \cite{ref08}. Measuring the spin squeezing relies on input and readout modes, constrained by their spatial overlap: the above result is valid for identical modes, whereas for different modes we find that the beam splitter coupling is altered by the modes' overlap, $\chi^{2}$, resulting in equivalent outcomes, though with a reduced coupling $r_{0}\rightarrow r_{0}\chi^{2}$ (Appendix B).

\emph{Spin squeezing analysis}--- Using Eq. (\ref{eq_Pu_HL}), we obtain the steady-state solution of $\hat{P}$, calculate its correlations via Eq. (\ref{eq_squeezed_correlations}), and find the spin squeezing parameter (\ref{eq_squeezing_param_def}),
\begin{eqnarray}\label{eq_squeezing_param}
\begin{aligned}
        \displaystyle
\xi^{2}&=1+2r_{0}\left(\mathcal{N}-\alpha_{\text{eff}}\sqrt{\mathcal{N}\left(\mathcal{N}+1\right)}\right)\rightarrow1-r_{0}+r_{0}\xi_{F}^{2}.
\end{aligned}
\end{eqnarray}
Here, $\alpha_{\text{eff}}=\alpha\frac{\left|r\right|}{r_{0}}$, with $r = r_{0}/(1+2i\frac{\delta-\Delta}{\Gamma}r_{0})$ being the array reflectivity \cite{ref15}, captures effective imperfections in the squeezed-vacuum source, such as phase-mismatch ($\delta\neq \Delta$) or impurity of squeezing ($\alpha<1$). The right-arrow signifies the result on resonance ($\delta = \Delta$), where a beam splitter transformation becomes apparent, exhibiting the weighted sum of input-field squeezed noise, $\xi_{F}^{2}$, and leakage-modes vacuum noise, $1$, with weights determined by the respective channel couplings $r_{0}$ and $1-r_{0}$.

\begin{figure}
  \centering
  \includegraphics[width=\columnwidth]{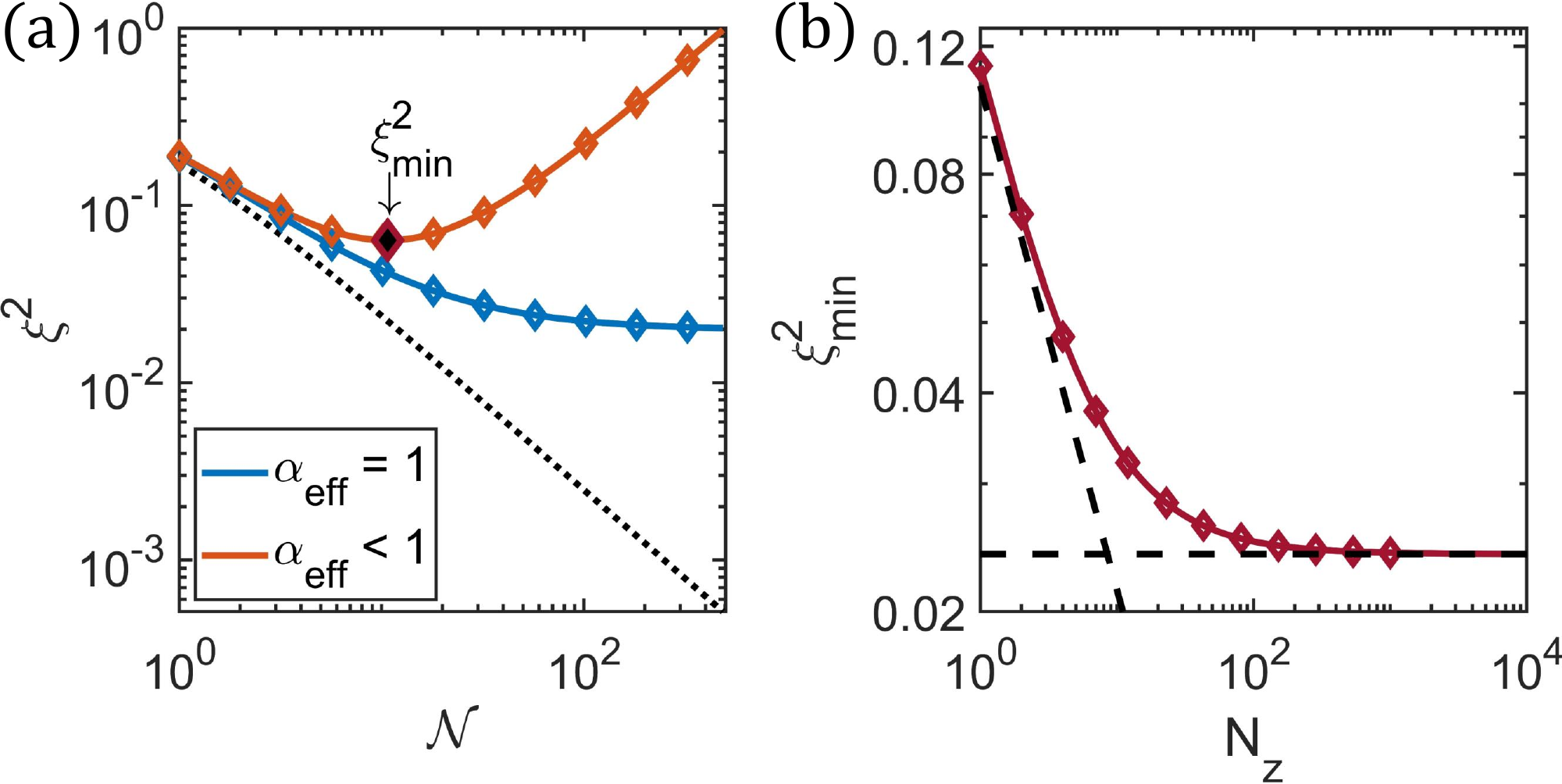}
  \caption{(a) Spin squeezing parameter vs. average number of incident photons, $\mathcal{N}$, calculated analytically from Eq. (\ref{eq_squeezing_param}) (solid lines) and numerically (diamond markers, see text for details). The dotted line represents the squeezing of the incident field, $\xi_{F}^{2}$. We consider two cases: $\alpha_{\text{eff}}=1$ and $\alpha_{\text{eff}}=0.999$, both for a resonant input field. On-resonance reflectivity $r_0$ is determined using Eqs. (\ref{eq_r0}) and (\ref{eq_emission_rates}) with an overlap parameter $\eta=0.99$, $N_z=10$ layers, and a non-collective decay rate $\gamma_{s}/\Gamma_{0}=0.1$. For the numerical calculations, we further consider the lattice spacings $a/\lambda=0.68$ and $a_{z}=\lambda$ to construct the full dipole-dipole interaction matrix between layers. (b) Minimal spin squeezing parameter, $\xi_{\text{min}}^{2}$, vs. the number of layers $N_{z}$ calculated analytically from Eq. (\ref{eq_optimal_squeezing}) (solid line) and numerically (diamond marker). The parameters match those in plot (a), but with $\alpha_{\text{eff}}=0.9999$. Dashed lines mark predicted asymptotic behaviors (see text).}\label{fig3}
\end{figure}

It is instructive to examine the dependence of the spin squeezing on system parameters. Fig. \ref{fig3}a shows the spin squeezing parameter as a function of the average photon number $\mathcal{N}$. The dotted line represents the ideal scenario, $r_{0}=1$ and $\alpha_{\text{eff}}=1$, where spin squeezing reaches its minimum value dictated by the input field, $\xi_{F}^{2}$, scaling as $1/\mathcal{N}$ for $\mathcal{N}\gg1$. Considering finite reflectivity $r_{0}<1$, we explore two scenarios: $\alpha_{\text{eff}}=1$ and $\alpha_{\text{eff}}<1$ representing effectively perfect and imperfect squeezing sources, respectively. For $\alpha_{\text{eff}}=1$, and for growing $\mathcal{N}$ (where $\xi_{F}^{2}\propto \mathcal{N}^{-1}\rightarrow 0)$, the spin squeezing is ultimately limited by the vacuum noise from the leakage modes, hence reaching the minimal asymptotic value $1-r_{0}$. This highlights the pivotal role played by the reflectivity $r_0$ in characterizing the transfer process.
When $\alpha_{\text{eff}}<1$, the competition between $\mathcal{N}$ and $\alpha_{\text{eff}}\sqrt{\mathcal{N}\left(\mathcal{N}+1\right)}$ in Eq. (\ref{eq_squeezing_param}) leads to a minimum point (Fig. \ref{fig3}a), found as
\begin{eqnarray}\label{eq_optimal_squeezing}
        \displaystyle
\xi_{\text{min}}^{2}=1-r_{0}+r_{0}\sqrt{1-\alpha_{\text{eff}}^{2}}.
\end{eqnarray}
The spin squeezing parameter is bound from below by the Heisenberg limit $1/(N^2 N_z)$ \cite{ref05}, motivating the study of the optimal spin squeezing $\xi_{\text{min}}^{2}$ for increasing number of layers $N_z$. This is plotted in Fig. \ref{fig3}b for the case where the squeezing is limited by $r_0$ (i.e. for good $\alpha_{\text{eff}}\lesssim 1$). We observe that while $\xi_{\text{min}}^{2}$ exhibits an initial decrease with $N_z$, it eventually reaches a limiting value. This behavior expresses two distinct asymptotic regimes originated in two corresponding loss mechanisms that contribute to $\gamma_{\mathrm{loss}}$ in Eq. (\ref{eq_emission_rates}): The non-collective scattering, $\gamma_{s}$, is independent of $N_{z}$, whereas leakage due to imperfect spatial overlap, $\left(1-\eta\right)\Gamma_{0}N_{z}$, increases linearly with $N_{z}$. For small $N_{z}$, $N_{z}\ll\gamma_{s}/\left(1-\eta\right)\Gamma_{0}$, $\gamma_{s}$ dominates and $\xi_{\text{min}}^{2}\simeq\frac{\gamma_{s}}{\Gamma_{0}}\frac{1}{N_{z}}$, leading to the initial decrease $\xi_{\text{min}}^{2}\propto N_z^{-1}$. Conversely, for large $N_{z}$, $\left(1-\eta\right)\Gamma_{0}N_{z}$ dominates, causing $r_{0}$ to reach its upper bound $\eta$, resulting in the limiting value $\xi_{\text{min}}^{2}\simeq1-\eta$. Hence, by increasing the number of layers $N_z$, spin squeezing substantially improves up to an asymptotic value determined by the spatial overlap $\eta$ between the input-light mode and the array.

To verify our analytical estimations, we performed numerical calculations on a model of interacting 2D layers derived from Eq. (\ref{eq_HL_position}) by relaxing our previous assumptions on the longitudinal positions of the layers [see (iii) and (iv) in the discussion below Eq. (\ref{trans})]. Within this description, collective dipoles of each 2D layer interact via an effective dipole-dipole interaction mediated by both propagating and evanescent modes, allowing to find the spin squeezing by numerically diagonalizing the resulting interlayer interaction matrix (Appendix D). Notably, we observe in Fig. \ref{fig3} excellent agreement between the analytical and numerical calculations, confirming the validity of our analytical approach within the regime of negligible evanescent-field interactions ($a< a_{z}$; see Appendix D for discussion on first order correction). Furthermore, we verified that  the calculations in all plots are consistent with our linearization assumption and the Heisenberg limit; in the case of high reflectivity, these entail  the condition $\mathcal{N}<N_{\text{eff}}$, where $N_{\text{eff}}=\eta\frac{2\pi w^{2}}{a^{2}}N_{z}$ is the effective number of atoms subjected to illumination.

\emph{3LAs scheme}--- So far, we have focused on an array of TLAs; however, our proposed methodology readily extends to arrays of 3LAs. For TLAs, squeezing is attained upon reaching a steady state thus requiring continuous illumination for sustaining squeezing; once illumination ceases, the state is damped by vacuum noise to a coherent spin state within a timescale $\Gamma^{-1}$. While the TLA scheme may be applicable to optical clocks using alkaline earth(-like) atoms \cite{ref30,ref44,ref45}, a more robust scheme is offered by considering a 3LA scheme. In 3LAs, squeezing can be transferred to the coherence between two ground atomic states, establishing a stable spin-squeezed state without relying on continuous illumination. For microwave clock transitions in alkali atoms such as Rb and Cs \cite{ref33,ref42,ref47}, this can lead to squeezing transfer to the clock transition, potentially enhancing clock operation accuracy. Our detailed analysis in Appendix C considers an atomic array composed of 3LAs with a stable level $\left|s\right\rangle$ in addition to the ground state $\left|g\right\rangle$ (Fig. \ref{fig2}b), focusing on their collective atomic spin $\hat{S}$ given by a weighted sum of $\hat{\sigma}_{\boldsymbol{n}_{\perp}n_{z}}^{gs}=\left(\left|g\right\rangle \left\langle s\right|\right)_{\boldsymbol{n}_{\perp}n_{z}}$ in analogy to the definition of $\hat{P}$ in (\ref{trans}). Since $\hat{P}$ mediates between the squeezed field $\hat{\mathcal{E}}$ and $\hat{S}$ via the coupling field $\Omega$, we can derive a dynamical equation for $\hat{S}$ by adiabatically eliminating $\hat{P}$, which adopts the form of Eq. (\ref{eq_Pu_HL}) with modified decay rates, shifts, and input fields (Appendix C). Consequently, in steady state we obtain a spin squeezing parameter for $\hat{S}$ that matches the form derived for TLAs in Eq. (\ref{eq_squeezing_param}). In particular, at two photon resonance, we reproduce the same results obtained for TLAs, with the individual decay $\gamma_s$ supplemented by the decay rate $\gamma_{\mathrm{se}}$ to $\left|s\right\rangle$ (Fig. \ref{fig2}b), which can be compensated for by adding layers (Fig. \ref{fig3}b). Importantly, the 3LA scheme entails the benefit of maintaining stable spin squeezing even when the fields $\Omega$ and $\hat{\mathcal{E}}$ are switched off, thereby increasing adaptability for various applications.

\emph{Discussion}--- Our work demonstrates the dissipative transfer of squeezed light into spin-squeezed states within an atomic array via strong light-matter interactions. The results highlight the key role of on-resonance reflectivity in quantifying quantum task efficiency, particularly in entanglement transfer; that is, optimizing the reflectivity emerges as a favorable strategy for an efficient spin-squeezing generation. Our approach holds potential for enhancing lattice-based optical atomic clocks, enabling them to operate beyond the standard quantum limit. In order to satisfy both the magic wavelength requirement for clock operation \cite{ref30} and the phase-matching condition  required during squeezing transfer ($a_{z}/\lambda\in\mathbb{N}$), the apparatus can integrate either adjustable layer spacing $a_z$ of a 3D optical lattice, or a tweezer array of 2D-lattice layers with interlayer spacing $a_z$ independent of the trapping magic wavelength. Going forward, the beam-splitter, 1D description of the presented technique for entanglement transfer, implies on its extended applicability. First, for the transfer of various quantum states beyond squeezing, and second, for its relevance to atomic systems beyond 3D-lattice arrays via their mapping to the 1D model \cite{ref20}.

\emph{acknowledgments}--- We acknowledge financial support from the Israel Science Foundation (ISF) grant No. 2258/20, the ISF and the Directorate for Defense Research and Development (DDR\&D) grant No. 3491/21, the Center for New Scientists at the Weizmann Institute of Science, the Council for Higher Education (Israel), and QUANTERA (PACE-IN). This research is made possible in part by the historic generosity of the Harold Perlman Family.

\appendix

\section{The incident field}
The free-evolving incident field projected onto the atomic dipole within the rotated frame is given by
\begin{eqnarray}\label{eq_EM_field_op}
\begin{aligned}
        \displaystyle
\hat{E}_{0}\left(\boldsymbol{r}\right)&=i\sum_{\boldsymbol{k}\mu}\sqrt{\frac{\hbar\omega_{\boldsymbol{k}}}{2\epsilon_{0}L^{3}}}\boldsymbol{e}_{\boldsymbol{k}\mu}\cdot\boldsymbol{e}_{d}^{*}\hat{a}_{\boldsymbol{k}\mu}(0)e^{i\boldsymbol{k}\cdot\boldsymbol{r}+i\left(\omega-\omega_{\boldsymbol{k}}\right)t},
\end{aligned}
\end{eqnarray}
where $\hat{a}_{\boldsymbol{k}\mu}(0)$ are operators in the Schr\"{o}dinger picture (at $t=0$). We divide the field into right- ($+$) and left- ($-$) propagating components $\hat{E}_{0,\pm}\left(\boldsymbol{r}\right)$ (containing only $k_z>0$ and $k_z<0$, respectively), and employ the projection onto the transverse input-mode from Eq. (\ref{Eu_transformation}), finding the input-field in the paraxial approximation as an expansion of 1D propagating waves \cite{ref20},
\begin{eqnarray}\label{eq_EM_field_1D}
        \displaystyle
{\cal E}_{\pm}\left(z\right)=i\sqrt{\frac{c}{L}}\sum_{k_{z}>0}\hat{a}_{uk_{z}}(0)e^{\pm ik_{z}z+i\left(\omega- c k_{z}\right)t},
\end{eqnarray}
with
\begin{eqnarray}\label{eq_au}
        \displaystyle
\hat{a}_{uk_{z}}=\frac{1}{L}\sum_{\boldsymbol{k}_{\perp}}\sum_{\mu}\boldsymbol{e}_{d}^{*}\cdot\boldsymbol{e}_{\boldsymbol{k}_{\perp}k_{z}\mu}\hat{a}_{\mu\boldsymbol{k}_{\perp}k_{z}}\tilde{u}\left(\boldsymbol{k}_{\perp}\right),
\end{eqnarray}
being the annihilation operator associated with the transverse mode $u\left(\boldsymbol{r}_{\perp}\right)$, and $\tilde{u}\left(\boldsymbol{k}_{\perp}\right)$ being the Fourier transform of $u\left(\boldsymbol{r}_{\perp}\right)$.
For a squeezed-vacuum input-field, the annihilation and creation operators satisfy \cite{ref39}
\begin{eqnarray}\label{eq_squeezed_correlations_a}
\begin{aligned}
        \displaystyle
\left\langle \hat{a}_{uk_{z}}^{\dagger}\hat{a}_{uq_{z}}\right\rangle  & =\delta_{k_{z}q_{z}}\left[N_{u}^{+}\left(k_{z}\right)+N_{u}^{-}\left(k_{z}\right)\right],\\
\left\langle \hat{a}_{uk_{z}}\hat{a}_{uq_{z}}\right\rangle  & =\delta_{2k-k_{z},q_{z}}M_{u}^{+}\left(k_{z}\right)+\delta_{2k+k_{z},-q_{z}}M_{u}^{-}\left(k_{z}\right).
\end{aligned}
\end{eqnarray}
Here $k=\omega/c$, and $N_{u}^{\pm}\left(k_{z}\right)$ and $M_{u}^{\pm}\left(k_{z}\right)$ are the squeezing spectrum parameters for the right (+) and left (-) propagating squeezed vacuum fields. Under the assumption of a paraxial squeezed vacuum reservoir and using the Markov approximation, we have calculated the correlations of the symmetric input field  as presented  in Eq. \ref{eq_squeezed_correlations}, with $\mathcal{M}=-\frac{1}{2}\left[M_{u}^{+}\left(\frac{\omega}{c}\right)+M_{u}^{-}\left(-\frac{\omega}{c}\right)\right]$ and $\mathcal{N}=\frac{1}{2}\left[N_{u}^{+}\left(\frac{\omega}{c}\right)+N_{u}^{-}\left(-\frac{\omega}{c}\right)\right]$.

\section{The spin squeezing parameter for a non-uniformly coupled system}
In this section, we outline the methodology for quantifying spin squeezing in non-uniformly coupled systems by generalizing the spin squeezing parameter defined by Wineland \emph{et al} \cite{ref06,ref07}.  While the conventional definition assumes a symmetric superposition of all spins relevant to uniform light-matter coupling, our system's non-uniform coupling renders it inapplicable. Nevertheless, for a non-uniformly coupled system with the majority of spins pointing in the same direction on the Bloch sphere, the system can be mapped to a system of uniformly coupled spins with a reduced effective atom-light coupling and a reduced effective atom number \cite{ref40}. Following the mapping procedure outlined in \cite{ref40}, we define effective spin operators ($i\in\{x,y,z\}$)
\begin{eqnarray}\label{eq_L}
        \displaystyle
\hat{L}_{i}=\frac{1}{f_{e}}\frac{1}{2}\sum_{\boldsymbol{n}_{\perp}n_{z}}f_{\boldsymbol{n}_{\perp}n_{z}}\hat{\sigma}_{\boldsymbol{n}_{\perp}n_{z}}^{i},
\end{eqnarray}
where the parameter $f_{e}$ is a normalization factor that ensures the effective spin operators satisfy the angular momentum commutation relation on average, $\left\langle \left[\hat{L}_{x},\hat{L}_{y}\right]\right\rangle =i\left\langle \hat{L}_{z}\right\rangle $, and is given by
\begin{eqnarray}\label{eq_f_e}
        \displaystyle
f_{e}=\frac{\sum_{\boldsymbol{n}_{\perp}n_{z}}f_{\boldsymbol{n}_{\perp}n_{z}}^{2}}{\sum_{\boldsymbol{n}_{\perp}n_{z}}f_{\boldsymbol{n}_{\perp}n_{z}}}\equiv\frac{\left\langle f^{2}\right\rangle }{\left\langle f\right\rangle }.
\end{eqnarray}
$f_{\boldsymbol{n}{\perp}n_{z}}$ characterizes the coupling strength between the atom $(\boldsymbol{n}{\perp},n_{z})$ and the electromagnetic field mode employed for phase measurement in the readout process. Based on the approach taken by Wineland \emph{et al}, the spin squeezing parameter is defined as the ratio of phase sensitivities in Ramsey measurements, however these sensitivities are now determined for effective spin operators, as described in the main text.

Considering that the spin squeezing parameter is defined for a measurement process, it is directly related to the readout optical mode; however, it is also impacted by the input mode which generates the squeezing, and that can in general differ from the readout mode. Consequently, when analyzing the spin squeezing parameter obtained from an experimental setup, two processes must be considered: generation and measurement of spin squeezing, which are associated with an input mode (generating) and a readout mode (measurement). We begin by analyzing the simple case in which both readout and input modes are identical, and then extend our analysis to encompass cases in which the two modes are mismatched.

When the readout and input modes are identical, the coupling between the dipoles and the electromagnetic modes can be expressed simply by the spatial profile of the input mode, resulting in $f_{\boldsymbol{n}_{\perp}n_{z}}=a u\left(\boldsymbol{r}_{\boldsymbol{n}_{\perp}}\right)\frac{1}{\sqrt{N_{z}}}e^{ik a_{z}n_{z}}$  (recalling $k=\omega/c$). We observe that the effective collective spin operator $\hat{L}_{-}$ is directly proportional to the collective dipole mode, such that $\hat{L}_{-}=\frac{\sqrt{\eta}}{f_{e}}\hat{P}$, where $\left\langle f^{2}\right\rangle =\eta$ (recalling $\eta$ as the spatial overlap between the mode and the array). The phase sensitivity of a coherent spin state (CSS), which is a product state of all spins in the same state, here $\left|g\right\rangle$, is given by $\text{Var}\left(\phi_{\mathrm{CSS}}\right)=\frac{f_{e}}{\left\langle f\right\rangle }$, resulting in a spin squeezing parameter from Eq. \ref{eq_squeezing_param_def}.

The next step is to determine the spin squeezing parameter in a broader context, considering a mismatch between the transverse components of the readout and the   mode. We assume that the modes are spatially matched in the $z$ direction (i.e. propagating waves), so we can divide the readout mode into its transverse and longitudinal components, $ f_{\boldsymbol{n}_{\perp}n_{z}}=\frac{1}{\sqrt{N_{z}}}e^{ik a_{z}n_{z}}f\left(\boldsymbol{r}_{\boldsymbol{n}_{\perp}}\right)$. The spin squeezing parameter is affected by the coupling between the readout mode and the collective dipole mode, $\hat{P}$, and, consequently, depends on the overlap between the readout mode, $f\left(\boldsymbol{r}_{\boldsymbol{n}_{\perp}}\right)$, and the input mode, $u\left(\boldsymbol{r}_{\perp}\right)$. The first step is to introduce a complete basis $\left\{u_{\alpha}\left(\boldsymbol{r}_{\perp}\right)\right\}_{\alpha}$ that spans the function space of the $xy$ plane, so that $\sum_{\alpha}u_{\alpha}\left(\boldsymbol{r}_{\perp}\right)u_{\alpha}^{*}\left(\boldsymbol{r}'_{\perp}\right)=\delta\left(\boldsymbol{r}_{\perp}-\boldsymbol{r}'_{\perp}\right)$. We select $u_{0}\left(\boldsymbol{r}_{\perp}\right)$ to be the input Gaussian mode $u\left(\boldsymbol{r}_{\perp}\right)$, with the orthonormal basis being the Hermite Gauss base. We define a set of collective dipole modes $\hat{P}_{\alpha}$ as
\begin{eqnarray}\label{eq_P_alpha}
        \displaystyle
\hat{P}_{\alpha}=\frac{a}{\sqrt{\eta}}\sum_{n_{z}=0}^{N_{z}-1}\sum_{\boldsymbol{n}_{\perp}}^{N^{2}}\frac{1}{\sqrt{N_{z}}}e^{ik a_{z}n_{z}}u_{\alpha}\left(\boldsymbol{r}_{\boldsymbol{n}_{\perp}}\right)\hat{\sigma}_{\boldsymbol{n_{\perp}}n_{z}},
\end{eqnarray}
where $\hat{P}_{0}=\hat{P}$ is the collective dipole mode that couples to the input squeezed vaccum mode. The modes $\hat{P}_{\alpha}$ satisfy the commutation relation
\begin{eqnarray}\label{eq_P_alpha_commute}
        \displaystyle
\left[\hat{P}_{\alpha},\hat{P}_{\alpha'}\right]=a^{2}\sum_{\boldsymbol{n}_{\perp}}^{N^{2}}u_{\alpha'}\left(\boldsymbol{r}_{\boldsymbol{n}_{\perp}}\right)u_{\alpha}\left(\boldsymbol{r}_{\boldsymbol{n}_{\perp}}\right)\equiv\eta_{\alpha\alpha'}.
\end{eqnarray}
For an infinite array with $w \gg a$, we can find that for the most dominant (paraxial) modes, the summation can be approximated as an integral and $\eta_{\alpha\alpha'} = \delta_{\alpha\alpha'}$. The function $f\left(\boldsymbol{r}_{\boldsymbol{n}_{\perp}}\right)$ can be represented in terms of $u_{\alpha}\left(\boldsymbol{r}_{\perp}\right)$,
\begin{eqnarray}\label{eq_f_alpha}
\begin{aligned}
        \displaystyle
\begin{array}{cc}
f\left(\boldsymbol{r}_{\boldsymbol{n}_{\perp}}\right)=&\sum_{\alpha}\sqrt{A_{\alpha}}u_{\alpha}\left(\boldsymbol{r}_{\perp}\right)f_{\alpha}\\f_{\alpha}\left(\boldsymbol{r}_{\perp}\right)=&\frac{1}{\sqrt{A_{\alpha}}}\intop d^{2}\boldsymbol{r}_{\perp}f\left(\boldsymbol{r}_{\perp}\right)u_{\alpha}\left(\boldsymbol{r}_{\perp}\right),
\end{array}
\end{aligned}
\end{eqnarray}
where $A_{\alpha}$  is a mode-related area scale. Accordingly, the effective spin operator can be expressed in terms of $\hat{P}_{\alpha}$,
\begin{eqnarray}\label{eq_L_alpha_base}
        \displaystyle
\hat{L}_{-}=\frac{1}{f_{e}}\frac{\sqrt{N_{z}\eta}}{a}\sum_{\alpha}\sqrt{A_{\alpha}}f_{\alpha}\hat{P}_{\alpha}.
\end{eqnarray}
In order to estimate the spin squeezing parameter, it is necessary to calculate the correlations between the various collective dipole modes $\left\langle \hat{P}_{\alpha}^{\dagger}\hat{P}_{\alpha'}\right\rangle$  and $\left\langle \hat{P}_{\alpha}\hat{P}_{\alpha'}\right\rangle$. For an infinite array, the Gaussian mode $\hat{P}_{0}$ is coupled to the squeezed vacuum, all the remaining modes are coupled to vacuum modes, and their correlations follow $\left\langle \hat{P}_{\alpha}\hat{P}_{\alpha'}^{\dagger}\right\rangle =\delta_{\alpha\alpha'}$, $\left\langle \hat{P}_{\alpha}\hat{P}_{\alpha'}\right\rangle=0$. Considering a finite array and finite beam, correlations are affected by the overlap of collective modes which leads to an emergence of correlations from undesired vacuum leakage modes (quantified by the decay rate $\left(1-\eta\right)\Gamma_{0}N_{z}$), which is proportional to the overlap between the Gaussian mode and leakage modes, $\left\langle \hat{P}_{\alpha}\hat{P}_{\alpha'}^{\dagger}\right\rangle \propto \eta_{0\alpha}$ , with $\eta_{0\alpha}$ defined in Eq. (\ref{eq_P_alpha_commute}). Our model is limited in its ability to characterize the dynamics of all collective modes due to its underlying paraxial approximation. Nevertheless, the overlap $\eta_{0\alpha}$ decays fast as the mode index increases, implying that only low indices which correspond to paraxial modes contribute to the correlations. As a result, the impact of non-paraxial modes becomes negligible, making it sufficient to calculate correlations exclusively for paraxial modes, whose dynamics are governed by our 1D model [Eq.(\ref{eq_Pu_HL})]. We calculate the correlations $\left\langle \hat{P}_{\alpha}\hat{P}_{\alpha'}\right\rangle =r\mathcal{M}\frac{\eta_{\alpha0}}{\eta}\frac{\eta_{\alpha'0}}{\eta}$ and $\left\langle \hat{P}_{\alpha}^{\dagger}\hat{P}_{\alpha'}\right\rangle =r_{0}\mathcal{N}\left(\frac{\eta_{\alpha0}}{\eta}\frac{\eta_{\alpha'0}}{\eta}\right)$ and find the spin squeezing parameter
\begin{eqnarray}\label{eq_squeezing_param_mismatch}
        \displaystyle
\xi^{2}=1+2r_{0}\chi^{2}\left(\mathcal{N}-\alpha_{\text{eff}}\sqrt{\mathcal{N}\left(\mathcal{N}+1\right)}\right)
\end{eqnarray}
with
\begin{eqnarray}\label{eq_chi}
        \displaystyle
\chi=\frac{\frac{1}{a}\intop d^{2}\boldsymbol{r}_{\perp}f\left(\boldsymbol{r}_{\perp}\right)u\left(\boldsymbol{r}_{\perp}\right)}{\sqrt{\left(a^{2}\sum_{\boldsymbol{n}_{\perp}}u^{2}\left(\boldsymbol{r}_{\boldsymbol{n}_{\perp}}\right)\right)\left(\sum_{\boldsymbol{n}_{\perp}}f^{2}\left(\boldsymbol{r}_{\boldsymbol{n}_{\perp}}\right)\right)}}
\end{eqnarray}
being the overlap parameter between the readout and input mode. In cases of a mismatch between the readout and input modes, the spin squeezing measurement is limited by the spatial overlap between the two modes, which can be characterized as a reduction in reflectivity from $r_{0}$ to $r_{0}\chi^{2}$.

\section{Three-level atomic array spin squeezing parameter}

Consider a 3LA comprised of a stable state $\left|s\right\rangle$ and a groung state $\left|g\right\rangle$, as depicted in Fig. (\ref{fig2})b. For this system, in addition to the collective operator $\hat{P}$, there exists another relevent collective atomic operator $\hat{S}$, which describes the dipole between the stable levels, $\hat{\sigma}_{\boldsymbol{n}_{\perp}n_{z}}^{gs}=\left(\left|g\right\rangle \left\langle s\right|\right)_{\boldsymbol{n}_{\perp}n_{z}}$, and is coupled to $\hat{P}$ by an external field, $\Omega$. Following the mapping procedure in \cite{ref20},  the dynamical equations of the 3LA are

\begin{eqnarray}\label{eq_3LA}
        \displaystyle
\begin{cases}
\dot{\hat{P}}= {} & \left[i\left(\delta-\Delta\right)-\frac{\Gamma+\gamma_{\text{loss}}}{2}\right]\hat{P}+i\Omega\hat{S}\\
&+i\sqrt{\gamma_{\text{loss}}}\hat{{\cal F}}(t)+i\sqrt{\Gamma}\hat{{\cal E}}(t)\\
\dot{\hat{S}}= & i\delta_{2}\hat{S}+i\Omega^{*}\hat{P}
\end{cases}
\end{eqnarray}
where $\delta_{2}=\delta-\delta_{c}$ is the detuning of the two photon transition from $\left|g\right\rangle$  to $\left|s\right\rangle$ and with $\delta_{c}$ being the detuning of the coupling field $\Omega$ from the $\left|s\right\rangle$ to $\left|e\right\rangle$ transition. The loss rate, $\gamma_{\text{loss}}$, may include an additional non-collective decay process, $\gamma_{se}$, which describes the decay from $\left|e\right\rangle$ to $\left|s\right\rangle$.
Adiabatically eliminating $\hat{P}$, we obtain the dynamical equation for $\hat{S}$,
\begin{eqnarray}\label{eq_S}
\begin{aligned}
        \displaystyle
\dot{\hat{S}}={} & \left(i\left(\delta_{2}-\Delta_{S}\right)-\frac{\gamma_{S,\text{loss}}}{2}-\frac{\Gamma_{S}}{2}\right)\hat{S}\\
&+i\sqrt{\gamma_{\text{loss}}}\hat{F}_{S}(t)+i\sqrt{\Gamma_{S}}\hat{{\cal E}}(t),
\end{aligned}
\end{eqnarray}
with
\begin{eqnarray}\label{eq_S_eff_param}
\begin{aligned}
        \displaystyle
\Gamma_{S}=\rho\Gamma,\ \gamma_{S,\text{loss}}=\rho\gamma_{\text{loss}},\\ \Delta_{S}=\rho\left(\delta-\Delta\right),\ \hat{F}_{S}=\sqrt{\rho}\hat{{\cal F}}
\end{aligned}
\end{eqnarray}
and
%
\begin{eqnarray}\label{eq_rho}
        \displaystyle
\rho=\frac{\left|\Omega\right|^{2}}{\left(\frac{\gamma_{\text{loss}}}{2}+\frac{\Gamma}{2}\right)^{2}+\left(\delta-\Delta\right)^{2}}.
\end{eqnarray}

The spin squeezing of the stable manifold of the 3LA is defined for the operator $\hat{S}$,
\begin{eqnarray}\label{eq_squeezing_param_S}
        \displaystyle
\xi^{2}=\min_{\theta}\text{Var}\left(\hat{S}e^{i\theta}+\hat{S}^{\dagger}e^{-i\theta}\right).
\end{eqnarray}
Eq. (\ref{eq_S}) for $\hat{S}$ has a structure identical to the dynamic equation for $\hat{P}$ [Eq. (\ref{eq_Pu_HL})]. As a result, the spin squeezing parameter follows a similar form, employing newly defined effective parameters. Accordingly, the spin squeezing parameter for the 3LA can be written as follows:
\begin{eqnarray}\label{eq_squeezing_param_3LA}
\begin{aligned}
        \displaystyle
\xi^{2}= {} & 1+2r_{0}\Biggl(\mathcal{N}\\
&-\alpha_{\text{eff}}\frac{1}{1-\frac{i\delta_{2}}{\frac{\Gamma_{S}}{2}+\frac{\gamma_{S,\text{loss}}}{2}+i\Delta_{S}}}\sqrt{\mathcal{N}\left(\mathcal{N}+1\right)}\Biggr).
\end{aligned}
\end{eqnarray}
For two-photon resonance, $\delta_{2}=0$, we obtain the exact form of Eq. (\ref{eq_squeezing_param}).

\section{Numerical calculations}
As part of the validation process of the analytical model, we conducted numerical calculations utilizing a model based on the paraxial approximation. This model allows for diverse values of $a_{z}$ and incorporates evanescent field interactions—both of which were constrained in the original analytical solution.
Following the methodology outlined in \cite{ref20}, we define a collective dipole for the $n_{z}$ layer,
\begin{eqnarray}\label{eq_Pnz_def}
        \displaystyle
\hat{P}_{n_{z}}=\frac{a}{\sqrt{\eta}}\sum_{\boldsymbol{n}_{\perp}}u\left(\boldsymbol{r}_{\boldsymbol{n}_{\perp}}\right)\hat{\sigma}_{\boldsymbol{n_{\perp}},n_{z}}.
\end{eqnarray}
Based on the paraxial approximation and assuming the array size is significantly greater than the dipole-dipole interaction lengthscale ($\lambda\ll Na$), Eq. (\ref{eq_HL_position}) can be transformed into a set of dynamical equations governing $\hat{P}_{n_{z}}$,
\begin{eqnarray}\label{eq_Pnz_EOM}
        \begin{aligned}
        \displaystyle
\dot{\hat{P}}_{n_{z}}={} & \left[i\left(\delta-\Delta\right)-\frac{\gamma_{s}}{2}-\frac{\Gamma_{0}}{2}\right]\hat{P}_{n_{z}}-\sum_{m_{z}\neq n_{z}}D_{n_{z}m_{z}}\hat{P}_{m_{z}}\\
&i\sqrt{\eta\Gamma_{0}}\hat{{\cal E}}_{n_{z}}\left(t\right)+\hat{F}_{u,n_{z}}+\hat{F}_{\eta,n_{z}},
\end{aligned}
\end{eqnarray}
Here, $\hat{{\cal E}}_{n_{z}}\left(t\right)=\frac{1}{\sqrt{2}}\left(\hat{{\cal E}}_{+}\left(0,t\right)e^{ika_{z}n_{z}}+\hat{{\cal E}}_{-}\left(0,t\right)e^{-ika_{z}n_{z}}\right)$ represents the paraxial input field on layer $n_{z}$. The noise term $\hat{F}_{u,n_z}$ is the transformation of $\hat{F}_{\boldsymbol{n}_{\perp}, n_z}$, and $\hat{F}_{\eta,n_{z}}$ accounts for the vacuum noise resulting from coupling to undesired leakage modes of layer $n_{z}$ (contributing to the overall leakage noise $F_{\eta}$). For a squeezed vacuum, using Eq. (\ref{eq_squeezed_correlations_a}), we derive the correlations of the layers' input fields:
\begin{eqnarray}\label{eq_E_nz_correlations}
\begin{aligned}
       \displaystyle
\langle \hat{{\cal E}}_{n_{z}}^{\dagger}(t)\hat{{\cal E}}_{m_{z}}(t')\rangle =&\mathcal{N}\cos\left(ka_{z}\left[n_{z}-m_{z}\right]\right)\delta\left(t'-t\right),\\
\langle \hat{{\cal E}}_{n_{z}}(t)\hat{{\cal E}}_{m_{z}}(t')\rangle =&\mathcal{M}\cos\left(ka_{z}\left[n_{z}+m_{z}\right]\right)\delta\left(t'-t\right),\\
\end{aligned}
\end{eqnarray}
and the nonzero correlations of the vacuum noise terms
\begin{eqnarray}\label{eq_vacuum_noise_correlations}
\begin{aligned}
       \displaystyle
\left\langle F_{u,n_{z}}\left(t\right)F_{u,m_{z}}^{\dagger}\left(t'\right)\right\rangle &=\gamma_{s}\delta_{m_{z}n_{z}}\delta\left(t'-t\right),\\
\left\langle \hat{F}_{\eta,n_{z}}\left(t\right)\hat{F}_{\eta,m_{z}}^{\dagger}\left(t'\right)\right\rangle &=\left(1-\eta\right)\Gamma_{0}\cos\left(ka_{z}\left[n_{z}-m_{z}\right]\right)\delta\left(t'-t\right).
\end{aligned}
\end{eqnarray}
The effective dipole-dipole interaction kernel $D_{n_{z}m_{z}}$ between the collective dipoles of different layers is given by:
\begin{eqnarray}\label{eq_D}
        \displaystyle
D_{n_{z}m_{z}}=\frac{\Gamma_{0}}{2}e^{ika_{z}\left|n_{z}-m_{z}\right|}+i\varepsilon_{n_{z}m_{z}},
\end{eqnarray}
with
\begin{eqnarray}\label{eq_D}
               \begin{aligned}
                       \displaystyle
\varepsilon_{n_{z}m_{z}}={} & \frac{\Gamma_{0}}{2}\sum_{\boldsymbol{m}_{\perp}\neq0}\frac{\left(\frac{\lambda}{a}\right)^{2}\left|\boldsymbol{m}_{\perp}\cdot\boldsymbol{d}\right|^{2}-1}{\sqrt{\left(\frac{\lambda}{a}\right)^{2}\left|\boldsymbol{m}_{\perp}\right|^{2}-1}}\\
&\times e^{-ka_{z}\left|n_{z}-m_{z}\right|\sqrt{\left(\frac{\lambda}{a}\right)^{2}\left|\boldsymbol{m}_{\perp}\right|^{2}-1}}.
\end{aligned}
\end{eqnarray}
The sum over $\boldsymbol{m}_{\perp}=\left(m_{x},m_{y}\right)$ accounts for contributions from the diffraction orders of each 2D layer \cite{ref15,ref20}. We can identify two distinct contributions to the interlayer effective dipole-dipole interaction. The first component accounts for interactions mediated by propagating fields, which determine the collective emission rate and induce energy shifts. Meanwhile, the second component, $\varepsilon$, elucidates interactions mediated through evanescent fields of higher diffraction orders. By diagonalizing the interaction kernel $D_{n_{z}m_{z}}$, we determine the steady state solution for $\hat{P}_{n_{z}}$. Subsequently, employing the transformation $\hat{P}=\frac{1}{\sqrt{N_{z}}}\sum_{n_{z}}e^{ika_{z}n_{z}}\hat{P}_{n_{z}}$ and using Eq. (\ref{eq_squeezing_param_def}), we can readily calculate the spin squeezing parameter.

The contribution of the evanescent fields diminishes exponentially within a range $\zeta_{\boldsymbol{m}_{\perp}}=a/\left(2\pi\sqrt{\left|\boldsymbol{m}_{\perp}\right|^{2}-\left(\frac{\lambda}{a}\right)^{2}}\right)$, which decreases as the diffraction order increases. Notably, when the interlayer distance exceeds $\zeta_{\left|\boldsymbol{m}_{\perp}\right|=1}$, $a_{z}>\zeta_{\left|\boldsymbol{m}_{\perp}\right|=1}$, $\varepsilon$ can be effectively treated as a perturbation. The results presented in this study (Fig. \ref{fig3}) pertain to a system operating within a domain where the impact of evanescent interlayer interaction is negligible. However, when $a$ approaches $a_{z}$, and $\zeta_{\left|\boldsymbol{m}_{\perp}\right|=1}$ becomes comparable to $a_{z}$, the contribution of the evanescent field becomes significant and cannot be disregarded, and a first-order correction must be applied. In \cite{ref20}, a first-order correction to the evanescent field was derived, resulting in an energy shift, $\Delta'$, given by
\begin{eqnarray}\label{eq_Delta_prime}
               \begin{aligned}
                       \displaystyle
\Delta'=\frac{1}{N_{z}}\sum_{n_{z}=0}^{N_{z}-1}\sum_{m_{z}\neq n_{z}}^{N_{z}-1}\varepsilon_{n_{z}m_{z}}e^{ika_{z}\left(n_{z}-m_{z}\right)}.
\end{aligned}
\end{eqnarray}
For example, in a 10-layer configuration with lattice spacings $a$ equal to $0.68a_{z}$, as illustrated in the main text, the first-order correction to the detuning amounts to $\Delta'=1.6\cdot10^{-4}\Gamma_{0}$, which is negligible, and therefore, the zeroth-order approximation suffices.  However, when $a$ is closer to $a_{z}$, setting $a$ to $0.95a_{z}$ instead, we observe that $\Delta' = 0.35\Gamma_{0}$, clearly indicating that the neglecting the evanescent interaction is not viable as $a$ approaches $a_z$.
\begin{figure}
  \centering
  \includegraphics[width=\columnwidth]{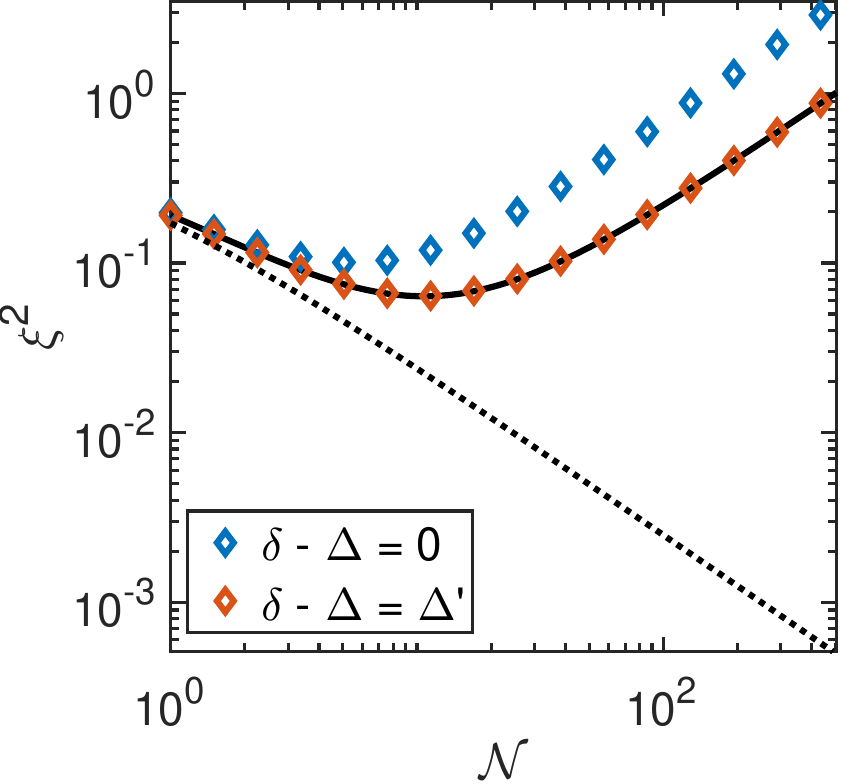}
  \caption{Spin squeezing parameter as a function of the average number of incident photons, $\mathcal{N}$, calculated analytically from Eq. (\ref{eq_squeezing_param}) on resonance $\delta-\Delta=0$  (solid lines). The dotted line represents the squeezed vacuum field quadrature $\xi_{F}^{2}$. The numerical spin squeezing parameter (diamond markers) is displayed for two detunings: $\delta-\Delta=0$ and $\delta-\Delta=\Delta'$. The plot corresponds to a 3D array with lattice spacings $a/\lambda=0.95$ and $a_{z}=\lambda$, $N_{z}=10$ layers, with an overlap parameter $\eta=0.99$, a non-collective decay rate $\gamma_{s}/\Gamma_{0}=0.1$ and attenuation $\alpha = 0.999$.}\label{fig4}
\end{figure}
To validate the accuracy of our analytical solution in the regime where $a$ approaches $a_{z}$, we conducted tests under a realistic experimental scenario with $\alpha\lesssim1$, and compared the analytical solution on-resonance with numerical calculations of the spin squeezing parameter that account for the evanescent field. The numerical calculations were performed for two cases including both zeroth-order and first-order corrections in $\varepsilon$,  which correspond to the choices of $\delta-\Delta=0$ and $\delta-\Delta=\Delta'$, respectively. The results depicted in  Fig. \ref{fig4} display the spin squeezing parameter plotted as a function of $\mathcal{N}$, with $a/a_{z}=0.95$, in a scenario where the first-order correction to the detuning holds significance. The on-resonance analytical solution does not agree with numerical calculations for the on-resonance case, $\delta-\Delta=0$; however, it does agree with numerical calculations for $\delta-\Delta=\Delta'$, which incorporate the first order correction to the energy shift. Clearly, relying solely on the zeroth-order treatment of the evanescent interaction is inadequate within this regime. Consequently, when working within parameter spaces characterized by large $a$, it becomes imperative to incorporate the first-order correction as it introduces a significant resonance shift. This correction can be seamlessly integrated into the analytical model, introducing an additional frequency shift, such that $\Delta\rightarrow\Delta+\Delta'$ as shown in \cite{ref20}.




\end{document}